\documentclass[doublecol]{epl2} 
\usepackage{mymacro}
\usepackage{subfigure}
\usepackage{amsmath}
\usepackage{color}
\usepackage{wrapfig}
\usepackage{cite}

\title{The Fredrickson-Andersen model with random pinning
on Bethe lattices and its MCT transitions}
\shorttitle{Fredrickson-Andersen model with random pinning
on Bethe lattices and its MCT transitions}

\author{Harukuni Ikeda\inst{1},  Kunimasa Miyazaki\inst{1} \and  Giulio Biroli\inst{2,3}}
\shortauthor{H. Ikeda \etal}

\institute{ \inst{1} Department of Physics, Nagoya University - Nagoya
  464-8602, Japan\\ \inst{2} IPhT, CEA/DSM-CNRS/URA 2306, CEA Saclay,
  F-91191 Gif-sur-Yvette Cedex, France. \\ \inst{3} Laboratoire de
  Physique Statistique, Ecole Normale Sup\'erieure, PSL Research
  University, 24 rue Lhomond, 75005 Paris, France. }

\abstract{ We investigate the dynamics of the randomly pinned
Fredrickson-Andersen model on the Bethe lattice. We find a line of
random pinning dynamical transitions whose dynamical critical properties
are in the same universality class of the $A_2$ and $A_3$ transitions of
Mode Coupling Theory. The $A_3$ behavior appears at the terminal point,
where the relaxation becomes logarithmic and the relaxation time
diverges exponentially. We explain the critical behavior in terms of
self-induced disorder and avalanches, strengthening the relationship
discussed in recent works between glassy dynamics and Random Field Ising
Model.}

\begin{document}

\maketitle
\section{Introduction}

It still remains a major challenge to fully understand
the origin of the dramatic slowing down of the dynamics of supercooled
liquids near the glass transition point with little, if any, sign of
structural orders.  Many supercooled liquids display rich and universal
dynamical behavior such as the two-step and non-exponential relaxation
of the correlation functions, the super-Arrhenius dependence of the
relaxation time, and spatially heterogeneous
dynamics\cite{debenedetti2001,Berthier2011,biroli2013}.

There are many theories attempting to describe the glass
transition\cite{Berthier2011,biroli2013}.  The mode coupling theory
(MCT) is very successful in describing
semi-quantitatively the time dependence of the density
correlation functions of the supercooled liquids at relatively high
temperatures\cite{bengtzelius1984,gotze2008,Berthier2011}.  However, it
is known that MCT fails to describe the dynamics at low
temperatures.  MCT predicts an artificial divergence of the
relaxation time well above the experimental glass transition
temperature\cite{bramb2009,Berthier2011}.

It is now considered that MCT is part of the random first order
transition (RFOT) theory, which is a thermodynamic theoretical approach
for the glass transition inspired by the similarity between supercooled
liquids and some mean-field spin glass
models\cite{PhysRevA.35.3072}.  MCT has the same
mathematical structure as that of the $p$-spin spherical model (PSM)
which is a mean-field model of the
RFOT\cite{PhysRevA.35.3072,kirkpatrick1989,bouchaud2004}.
Curiously, a different theory also predicts MCT-like dynamics at the
mean-field level. The dynamical facilitation scenario (DF) claims that
the glass transition is a purely dynamical transition {\it without} any
thermodynamic singularity \cite{chandler2009,biroli2013}.  
DF is based on the kinetically constrained models (KCM), which have
trivial thermodynamic properties but show complex glassy slow
dynamics\cite{chandler2009,ritort2003}.  The Fredrickson-Andersen model
(FA), as well as other KCMs, has been shown to
display the same scaling law of that of MCT in the mean-field
limit\cite{fredrickson1984,sellitto2005,sellitto2015}.

The fact that completely different theories (one is thermodynamic and
the other is kinetic) predict similar dynamics---and similar to MCT---in
the mean-field limit suggests that there is an underlying universality
hidden in MCT.  This was indeed already discussed in
\cite{andreanov2009}, where the MCT scaling laws were obtained by a
Landau-like expansion.  Recently, it was shown that the MCT criticality
is related to the one of the Random Field Ising Model (RFIM)
\cite{franz2011,franz2013r} 
and Franz and Sellitto have shown that the finite size scaling of the
critical dynamics of the FA model on the Bethe lattice are indeed consistent
with that of the RFIM\cite{franz2013}.

In this work, we investigate thoroughly the universal structure of MCT by
focusing on more general cases (always in the KCM context), studying
whether the relationship with MCT still holds, and unveiling its physical
content.  In order to do so, we take advantage of recent results on the
glass transition of randomly pinned systems
\cite{kim2003,krakoviack2005,karmakar2012,jack2012random,cammarota2012,kob2013,ozawa2015,ikeda2015},
a fluid where a fraction of constituent particles are frozen or pinned.
Theoretical analysis
\cite{cammarota2012,cammarota2013} predicts that by pinning a fraction
$c$ of particles from an equilibrium configuration at temperature $T$,
the glass transition temperature $T(c)$ rises until it reaches a
terminal critical point where it ends.  The properties of the glass
transition remain the same along the line but change at the terminal
point.  MCT predicts several anomalous dynamical behaviors for randomly
pinned systems\cite{krakoviack2005,cammarota2012aging,szamel2013}.  It
predicts an MCT critical line where the transition remains, in the MCT
terminology, of $A_2$ type, until the terminal point is reached where it
becomes $A_3$ type.  The dynamical behavior at the $A_3$ transition is
qualitatively different.  For example, the correlation function exhibits
single logarithmic decay\cite{gotze1989}, instead of the usual two-step
relaxation \cite{gotze2008}.  The relaxation time increases
exponentially toward the $A_3$ transition point, while it increases
algebraically toward the $A_2$ transition point\cite{gotze1989}.
Moreover, the critical behavior at the terminal point was shown to be
related to the critical behavior of the RFIM at its continuous
transition along the hysteresis line
\cite{cammarota2013,cammarota2012aging,franz2013r,nandi2014critical} and
not at its spinodal transition.
The randomly pinned FA model thus provides a very useful setting to
analyze the universal structure of MCT. Our aim will be on the one hand
to check that the properties of the dynamical transition do not vary
along the critical line induced by pinning and coincide with the ones
predicted by MCT for the $A_2$ singularity and on the other hand that
they do change at the terminal point, where instead they becomes the one
predicted by MCT for the $A_3$ singularity.  We shall also analyze the
relationship with the RFIM and work out the physical mechanism behind
it.  In the a previous study, two of us have already investigated the
static properties of the FA model with random pinning and indeed found
some evidences of the scenario presented above \cite{ikeda2015}.

\section{Model and phase diagram}
\label{184221_22May16}
\begin{figure}
 \onefigure[width=8.5cm]{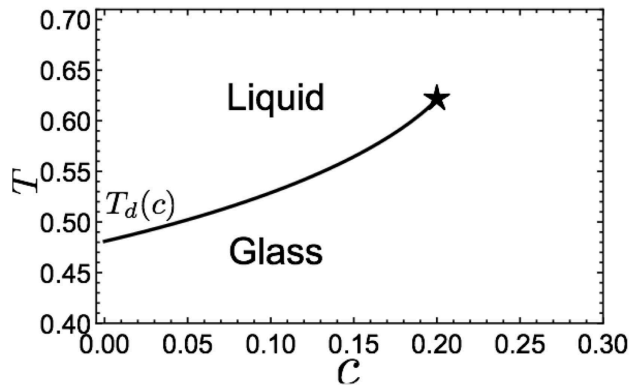} \caption{\small{The phase diagram of
 the model. The solid line denotes the putative $A_2$ transition line, $T_d(c)$.
 The filled star denotes the putative $A_3$ transition point, $c_c=0.2$ and
 $T_c=T_d(c_c)\approx 0.621$. }} \label{phase}
\end{figure}
We consider the FA model with random pinning on the regular random graph
with connectivity $z=k+1=4$\cite{ikeda2015}.  The Hamiltonian of the
model is given by $H = -\frac{1}{2}\sum_{i=1}^N \sigma_i$, where
$\sigma_i\in \{-1,+1\}$ denotes the binary spin variable on the $i$-th
site. We pick up a fraction (denoted as $c$) of spins randomly from $N$
spins and ``pin'' them.  The pinned spins are not allowed to move from
their initial equilibrium configuration. The time evolution rule of the
model is the following\cite{sellitto2005}. We randomly select a spin and
flip it with the probability $w(\sigma_i\to
-\sigma_i)=\min\{1,e^{-\sigma_i/T}\}$ if the spin is {\it not} pinned
and there are more than $f=2$ number of downward spins in its neighbor.
Otherwise, we do not flip the spin\cite{ikeda2015}. We use $N$
iterations of this process as a unit of time.  To characterize the slow
dynamics, we observe the persistence function, $\phi(t)$, which is the
fraction of the unflipped spins in the time span $[0,t]$. The analysis
of the long time limit of the persistence function,
$\phi=\lim_{t\to\infty}\phi(t)$, is particularly
simple\cite{sellitto2005}, since it can be mapped into the bootstrap
percolation (BP) problem\cite{chalupa1979}.  In the previous paper, we
have calculated $\phi$ analytically as a function of $T$ and
$c$\cite{ikeda2015}. The phase diagram from $\phi$ is shown in
Fig.~\ref{phase}.  The solid line, $T_d(c)$, is the transition line.
When $T>T_d(c)$, the system is in the ``liquid phase'' characterized by
the low $\phi$ value and when $T>T_d(c)$, the system is in the ``glass
phase'' characterized by the high $\phi$ value.  $T_d(c)$ rises with
increasing $c$ and terminates at $(c_c,T_c)\approx(0.2,0.621)$.
Approaching $T_d(c)$ for $c<c_c$, $\phi$ changes
discontinuously\cite{ikeda2015} as it would happen at an $A_2$
transition in the MCT terminology. Just below $T_d(c)$, $\phi$ changes
as $\phi-\phi_d(c) \propto (T-T_d(c))^{1/2}$, where $\phi_d(c)$ is the
fraction of the frozen spin at the transition temperature.  On the other
hand, at $(c_c,T_c)$, $\phi$ behaves differently as $\phi-\phi_d(c_c)
\propto (T-T_c)^{1/3}$, where $\phi$ changes continuously but singulary
as MCT predicts for the $A_3$ transition\cite{ikeda2015}.  Note that
$\phi$ only contains information about the long time limit. It is still
unclear how the above singularities of $\phi$ affect the explicit time
dependent dynamical quantities of the system, especially how the
dynamics is altered at the terminal point.  Below, we show that those
dynamical quantities indeed show the singular behaviors characteristic
of $A_2$ and $A_3$ transitions.

 \section{Persistence Function and Critical Dynamic Scaling}
\begin{figure}
 \onefigure[width=8.5cm]{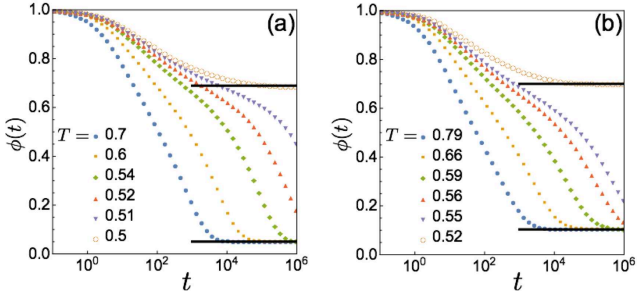} \caption{\small{Persistence
 functions approaching the random pinning transition (not at the
 terminal point). (a) The results for $c=0.05$. The filled symbols
 denote the results of numerical simulation where the average is taken over
 $10$ independent samples. The solid lines denote the fraction of the
 frozen spins estimated by the analytical equation. $\phi(t)$ converges
 to the value the fraction of the frozen spins in the long time
 limit. (b) The same data for $c=0.1$.}}  \label{finitec}
\end{figure}

Hereafter, we investigate the dynamical properties of this model in more
detail. We perform the Monte Carlo simulation (MC) by the
faster-than-clock algorithm\cite{krauth2006}.  The systems size is
$N=2^{18}$ unless specifically mentioned, which is large enough to
neglect the finite size effects. In this section, we focus on the time
dependence of the persistence function, $\phi(t)$, and vary the
temperature at fixed $c$.  First, we present the results far from the
terminal point. In this case, the dynamics should be of the standard
$A_2$ type. This is indeed the case, as we show in
Fig.~\ref{finitec} for $c=0.05$ and $c=0.1$.  The transition
temperatures are $T_d(c=0.05)\approx 0.502$ and $T_d(c=0.1)\approx
0.529$, respectively. As we approach $T_d(c)$ from above, the relaxation
time increases and $\phi(t)$ develops a plateau.  This two-step
relaxation behavior is commonly observed for many glassy materials
including KCMs on the Bethe lattice {\it without} random pinning
($c=0$)\cite{sellitto2005,arenzon2012,sellitto2013}.  In the long time
limit, $\phi(t)$ converges to a constant value, $\phi(\infty)$, which
coincides with $\phi$ as shown in Fig.~\ref{finitec}.  The discontinuous
jump of $\phi$ at the transition is directly connected to the two-step
relaxation of $\phi(t)$.

\begin{figure}
 \onefigure[width=8.5cm]{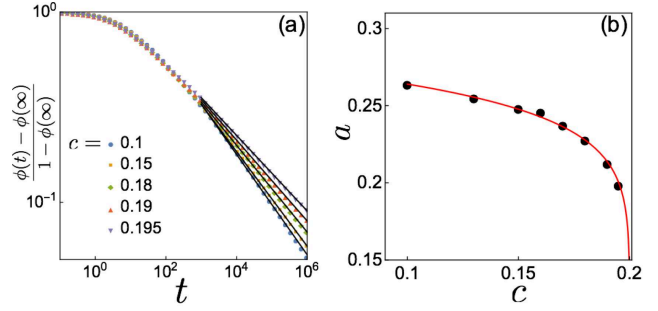} \caption{\small{ (a) The persistence
 function at the putative $A_2$ transition points, $T=T_d(c)$. The
 filled symbols are results of the numerical simulation where the system
 size is $N=2^{19}$ and the average is taken over $10^2$ independent
 samples. The solid lines are results of power law fitting.  (b) The $c$
 dependence of the critical exponent $a$. The filled cycles are data
 obtained by the numerical simulation.  The solid line is the power law
 fitting which goes to zero at $c=0.2$.}}  \label{fig1}
\end{figure}

At the transition temperature, $T=T_d(c)$, the relaxation of
$\phi(t)$ toward $\phi(\infty)$ is well fitted by a power law as expected from MCT;
\begin{align}
 \frac{\phi(t)-\phi(\infty)}{1-\phi(\infty)}=B t^{-a}.\label{144933_6May16}
\end{align}
In
Fig.~\ref{fig1} (a), we show the results obtained by fitting the results of our numerical
simulations for several $c$'s, where the filled symbols denote the
numerical results and the solid lines denote the fits by
eq.~(\ref{144933_6May16}).
The values of $a$ for various $c$'s estimated by the numerical
simulations are plotted in Fig.~\ref{fig1} (b). The exponent $a$ decreases sharply as
$c$ approaches the terminal point of the critical line, {\it i.e.}, the putative $A_3$ transition, at
$c_c=0.2$. A qualitatively similar result was obtained for the
multi-component extension of the FA model which also exhibits the $A_3$
transition\cite{arenzon2012}.

The sharp decrease of $a$ is consistent with the result of MCT which
predicts that $a$ decreases and eventually vanishes on the $A_3$
transition point\cite{gotze1989}.  The vanishing behavior of the
critical exponent suggests that eq.~(\ref{144933_6May16}) should be
replaced by a different scaling function at the $A_3$ transition point.
MCT predicts that $\phi(t)$ near the $A_3$ transition point follows
the scaling law\cite{gotze1989}:
\begin{align}
\frac{\phi(t)-\phi_c}{1-\phi_c} = (T-T_c)^{1/3} f\left((T-T_c)^{1/6}\log(t) \right),
\end{align}
where $\phi_c$ denotes $\phi$ at the $A_3$ transition point,
$(c_c,T_c)$.  The scaling function $f(x)$ behaves as $x^{-2}$ at small
argument and linearly (as $ax+b$) at large argument.  In order to test this scaling
law, we focus on the large $x$ regime\footnote{In order to test the
small $x$ regime, one would need times much larger than the ones
available in our simulations.} in which one should find
\begin{align}
\frac{\phi(t)-\phi_c}{1-\phi_c} = -B\log(t/\tau_\beta)\label{222037_6May16}
\end{align}
with the prefactor $B$ and the relaxation time\footnote{We denote the
relaxation time $\tau_\beta$ for there are not two distinct regimes at
the $A_3$ critical point but just one, that we denote $\beta$.}
$\tau_\beta$ scaling as
\begin{align}
B\propto (T-T_c)^{1/2},\ \ \log\tau_\beta\propto (T-T_c)^{-1/6}.\label{171523_3Jun16}
\end{align}

\begin{figure}
 \onefigure[width=8.5cm]{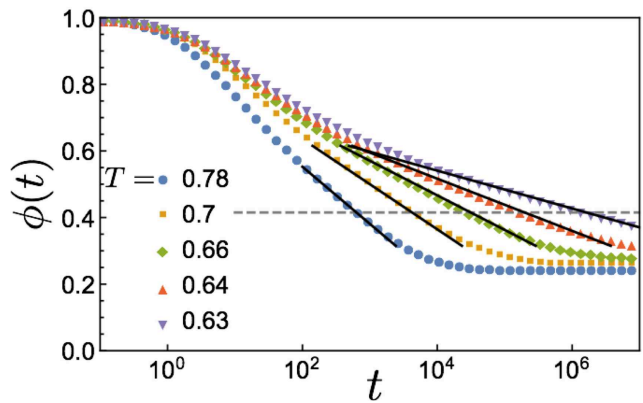} \caption{\small{The persistence
 functions near the terminal point of the random pinning transition
 line. The value of the pinned density is fixed at $c=c_c=0.2$. The
 filled symbols are simulation results where the average is taken over $10^2$
 independent samples.  The solid lines are logarithmic fit. The
 horizontal dashed line denotes $\phi_d(c_c)$.  }} \label{fig2_1}
\end{figure}
\begin{figure}
 \onefigure[width=8.5cm]{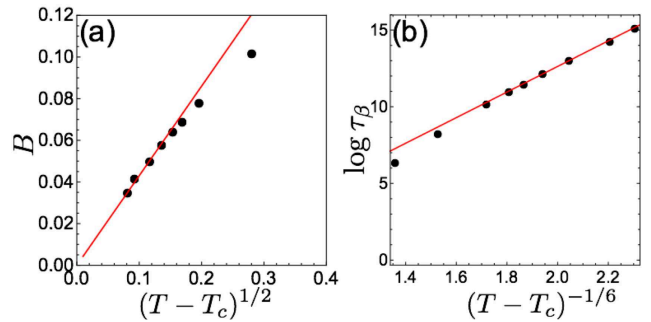} \caption{\small{ (a) The temperature
 dependence of $B$ near the putative $A_3$ transition temperature $T_c$.
 The value of pinned density is fixed at $c=c_c=0.2$. The filled cycles
 are results of the numerical simulations where the average is taken over
 $10^2$ independent samples.  The solid lines indicate the MCT
 prediction, $B\propto (T-T_c)^{1/2}$.  (b) The temperature dependence
 of $\tau_\beta$ near the $A_3$ transition point.  The filled symbols
 represent the numerical results and the solid line represents the MCT
 prediction, $\log\tau_\beta\propto (T-T_c)^{-1/6}$.  }} \label{fig2_2}
\end{figure}

In order to confirm eq.~(\ref{222037_6May16}), we fix $c=c_c=0.2$ and
calculate $\phi(t)$ for several temperatures slightly above the $A_3$
transition temperature, $T_c\approx 0.621$.  The results are shown in
Fig.~\ref{fig2_1}.  Near the $A_3$ transition point, $\phi(t)$ shows
single decay instead of the two-step relaxation as shown in
Fig.~\ref{fig2_1}.  This is consistent with the continuous change of
$\phi$ at the $A_3$ transition point, as reported in our previous
work\cite{ikeda2015}. In the intermediate time scale where $\phi(t)$ is
very close to $\phi_c$ and the scaling variable $x$ is large, $\phi(t)$
is well fitted by the MCT scaling law, eq.~(\ref{222037_6May16}), as
shown in Fig.~\ref{fig2_1}.  $B$ and $\tau_\beta$ obtained by the
fitting are shown in Fig.~\ref{fig2_2} with the scaling law,
eq.~(\ref{171523_3Jun16}), predicted by MCT.  One can see that near the
$A_3$ transition temperature, $B$ and $\tau_\beta$ indeed follow the MCT
scaling law.  Our results clearly support that the critical dynamic
scaling at the terminal point of the random pinning transition line for
the FA model is the one predicted by MCT at the $A_3$ dynamical
transition.

\section{Critical Fluctuations}
\begin{figure}
\onefigure[width=8.5cm]{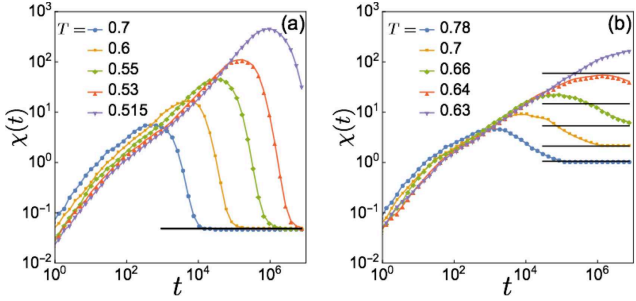} \caption{\small{(a) Susceptibilities
 near the putative $A_2$ transition point. The pinned density is fixed
 at $c=0.05$.  The filled symbols denote the numerical results where the
 average is taken over $10^2$ independent samples. The solid lines denote
 the values estimated by the bootstrap percolation process (see
 text). (b) The same date near the putative $A_3$ transition point. The
 pinned density is fixed at $c=c_c=0.2$.}} \label{fig6}
\end{figure}

We now focus on the behavior of dynamic fluctuations that we expect to
also become critical at the transition. In particular, we
analyze
\begin{align}
 \chi(t) \equiv
 N \left[\left\langle\phi(t)^2\right\rangle-\left\langle\phi(t)\right\rangle^2\right], 
\end{align}
where the bracket denotes the average for both the initial conditions
and thermal noises\cite{berthier2011overview}.  We calculate $\chi(t)$
approaching a point on the random pinning transition line ($c=0.05$) and
approaching the terminal point ($c_c=0.2$). The results are shown in
Fig.~\ref{fig6} (a) and (b);
$\chi(t)$ first increases with $t$ and reaches the maximum, $\chi^*$, at
$t=t^*$. At large times $\chi(t)$ decreases and converges to constant
values in the long time limit, $\lim_{t\to\infty}\chi(t)=\chi_{\infty}$.
Note that the values of $\chi_{\infty}$ coincide with the fluctuations
of the fraction of frozen spins, $\phi$, which can be easily calculated
by the numerical simulation of the bootstrap percolation
(BP). $\chi_{\infty}$ estimated by the BP are shown by solid lines in
Fig.~\ref{fig6} (a) and (b).

\begin{figure}
\onefigure[width=8.5cm]{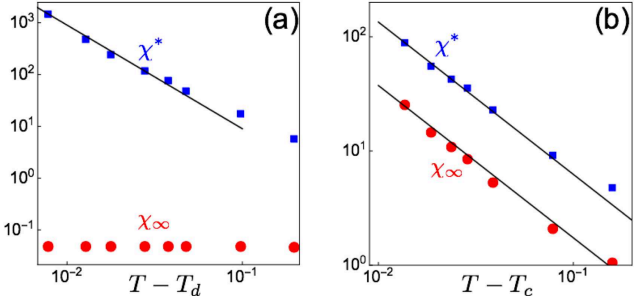} \caption{ \small{ (a) The peak value
 of the susceptibilities, $\chi^*$, and the values of the
 susceptibilities in the long time limit, $\chi_{\infty}$ near the putative $A_2$
 transition temperature. The pinned density is fixed at $c=0.05$.  The
 filled symbols are results obtained by the numerical simulation.  The
 solid line represent the power law scaling, $\chi^*\propto
 (T-T_d)^{-2}$.  (b) The same date near the putative $A_3$ transition
 temperature. The pinned density is fixed at $c=c_c=0.2$. The filled
 symbols represent the results obtained by the numerical simulations.
 The solid line represent the power law scaling, $\chi^*\propto
 \chi_{\infty}\propto (T-T_c)^{-4/3}$. }} \label{fig7}
\end{figure}

MCT provides detailed predictions regarding dynamical
fluctuation\cite{franz2000,biroli2006,berthier2007,franz2011}. In the
$A_2$ case, $\chi(t)$ varies on a time-scale of the order of the
relaxation time, displays a diverging peak $\chi^*\propto (T-T_d)^{-2}$,
and a featureless long-time limit $\chi_{\infty}$.
This is indeed what we find in our numerical simulations, see
Fig.~\ref{fig6} (a) and Fig.~\ref{fig7} (a),
in agreement with recent results for $c=0$~\cite{de2016}.

The MCT predictions for the $A_3$ case are qualitatively different
\cite{nandi2014critical}\footnote{In \cite{nandi2014critical} the IMCT
susceptibility was studied, $\chi(t)$ scales as its square as explained
in \cite{berthier2007,franz2011}.  }:
\begin{align}
 \chi(t)= \abs{T-T_c}^{-4/3} g\left((T-T_c)^{1/6}\log(t) \right),
 \label{scaling2}
\end{align}
where the scaling function $g(x)$ tends to a constant at large argument
hence implying that both $\chi^*$ and $\chi_{\infty}$ diverge as
$\abs{T-T_c}^{-4/3}$.  Again, this is what we find in our numerical
simulations, see Fig.~\ref{fig6} (b) and Fig.~\ref{fig7} (b) where we
fit the numerical data, as shown by the solid lines in Fig.~\ref{fig7}
(b). The agreement is very good, signalling that MCT predictions
\cite{nandi2014critical} hold also for the critical behavior of the
dynamical fluctuations at the terminal point of the random pinning
transition line.

\section{Self-induced Disorder and Avalanches}
We now show that the criticality found at the terminal point in the FA
model is related to self-induced disorder and to the continuous
transition along the hysteresis line of the Random Field Ising Model
(RFIM) and its avalanches. This is indeed expected since this
relationship was shown recently to hold for MCT at the $A_3$ transitions
\cite{franz2011,cammarota2013,nandi2014critical,nandi2016}.  Since the
critical behavior is presented in both $\chi^*$ and $\chi_\infty$, we
focus on the latter which can be analyzed using BP techniques
\cite{branco1993probabilistic,sellitto2005}. The fluctuation of $\phi$,
which leads to $\chi_\infty$, is due to different initial conditions
which play the role of the different realization of the quenched
randomness\cite{franz2011}. For instance, the initial fraction of the
upward spins, $p$, fluctuates for different initial conditions. This
causes sample-to-sample fluctuation, $\delta \phi\propto (\partial
\phi/\partial p)\delta p$, and leads to a contribution to the
susceptibility which leads $\chi_\infty\propto (\partial \phi/\partial
p)^2\left\langle\delta p^2\right\rangle\propto (\partial \phi/\partial
p)^2$. Near the transition point, $p$ can be expressed as a linear
function of $T-T_c$ and the derivative by $p$ can be replaced by $T$.
Thus, we obtain $\chi_{\infty}\propto
\left(\partial\phi(T_c)/\partial T\right)^2 \propto \abs{T-T_c}^{-4/3}$, which
shows the same divergence found for the RFIM at the continuous
transition along the hysteresis line.  This is not a coincidence;
actually all the critical mean-field behavior of the $A_3$ dynamical
transitions, and hence of the terminal point, is the same (see
\cite{nandi2014critical} for a detailed comparison) \footnote{Actually
in \cite{nandi2014critical} the comparison was done with the equilibrium
continuous RFIM transition which has at the mean-field level the same
critical behavior. Following \cite{nandi2016} we think that the
continuous transition along the hysteresis line is a better comparison
from the phenomenological point of view in particular because avalanches
appear explicitely.}.  The previous arguments show that the dynamical
critical behavior is produced by the combination of self-induced
disorder and the singular dependence of $\phi \sim |T-T_c|^{1/3}$.  Let
us now unveil that the physical mechanism behind this singular
dependence are avalanches identical to the ones present at the
continuous transition along the hysteresis line of the RFIM.  The key
idea is that by changing the temperature, one changes the fraction of,
say, up spins but this in turn leads to a much larger change of blocked
spin. The reason is precisely avalanches: by increasing of a factor of
two the distance ($\varepsilon=|T-T_c|$) from the transition, a fraction
of the order of $\varepsilon$ of spins becomes suddenly unblocked
because their number of pointing down nearest neighbours becomes larger
than $f=2$. This however leads to a cascade process since some
neighbours of the unblocked spins become unblocked and so on and so
forth. Thus by changing $\varepsilon$ by a factor of two, $\phi$
decreases by roughly $\varepsilon \langle S \rangle$ where $S$ is the
size of the avalanche, {\it i.e.}, the cascade process discussed above
created by unblocking a random spin. By generalizing the computation
performed for BP \cite{shukla2008} to the $c>0$ case, we have obtained
the distribution function of the size of the avalanche which acquires a
scaling form close to the transition at the terminal
point\cite{ikeda2016}:
  \begin{align}
   P(S) &= \frac{1}{S^{\tau}} h\left(S|T-T_c|^{4/3} \right),
\end{align}
where $\tau=3/2$ and the scaling function $h(x)$ coincides with the one
computed for the RFIM at the continuous transition along the hysteresis
line\cite{PhysRevLett.70.3347} thus strengthening the relationship
discussed above \footnote{We repeated the computation for
transitions in the $A_2$ universality class and found as expected an
avalanche distribution that coincides with the one obtained for the
spinodal of the RFIM\cite{PhysRevB.53.14872}.}. Using this result, one
finds that the average avalanche size scales as $\varepsilon ^{-2/3}$
thus leading to $\phi \sim \varepsilon^{1/3}$ and providing the final
missing piece to explain the critical behavior at the terminal point.
In conclusion, although the explicit time-dependence cannot be obtained
in this way, the critical behavior is fully understood even
quantitatively in terms of self-induced disorder and avalanches.
\section{Summary and discussion}
We investigated the equilibrium dynamics of the Fredrickson-Andersen
model on the Bethe lattice with random pinning as a mean-field model of
the dynamical facilitation scenario. We found a dynamical phase diagram 
qualitatively identical to the one predicted for RFOT (it would be 
interesting to check whether also the predictions for the aging dynamics hold)  \cite{cammarota2012aging}.
We showed that the persistence
function, $\phi(t)$, and the dynamical fluctuations, $\chi(t)$, display a
critical behavior consistent with the prediction of the mode coupling
theory (MCT) for both the $A_2$ and $A_3$ transitions.
We also showed that the critical behavior at the terminal point is
tightly related to the one of the RFIM and explicitly explained the
origin of this relationship in terms of self-induced disorder and
avalanches.

Our work fully exposes the universal character of MCT as a generic
mean-field dynamical transition. It appears not only in models
characterized by rugged energy landscapes but also in models with
trivial thermodynamics but glassy dynamics. The predictive power in
terms of scaling laws at $A_2$ and $A_3$ singularity for the FA model is
a remarkable sign that MCT indeed holds also for cooperative KCMs at the
mean-field level.  In finite dimensions however, fluctuations on top of
the mean-field theory are expected to lead to drastic changes
\cite{nandi2016,rizzo2014}.  Actually, in this case, very different
physical mechanisms are expected to be at play for KCMs and systems
characterized by rugged energy landscapes respectively.  More work is
needed---and certainly worth doing--- to understand the role of
fluctuations and their effect on the mean-field theory.

The other important result of our work is to show explicitly the role of
avalanches in determining the MCT critical behavior in agreement with
recent results \cite{nandi2016}.  Although the analysis of avalanches
does not allow to determine also time-dependent quantities, 
 the distribution functions of the mobile region for deep supercooled
liquids do show similar power law like behaviors
\cite{donati1999,weeks2000,gebremichael2004} and sign of avalanche
motion in glassy dynamics was found in experiments
\cite{PhysRevLett.102.088001} and in simulations
\cite{PhysRevLett.105.135702}.  This is certainly a topic worth further
studies both for its theoretical and phenomenological relevance.

\acknowledgments We thank C. Cammarota for feedback and comments on this work. 
H. I. and K. M acknowledge JSPS KAKENHI Grant Number
JP16H04034, JP25103005, JP25000002, and the JSPS Core-to-Core program.
H. I. was supported by Program for Leading Graduate Schools
``Integrative Graduate Education and Research in Green Natural
Sciences'', MEXT, Japan and JSPS Grant Number JP16J00389.
G. B. acknowledges support from the ERC grants NPRGGLASS and by a grant
from the Simons Foundation (\#454935, Giulio Biroli)

\section{Appendix}

Here we derive the avalanche distribution of the BP with random pinning
on the $z=k+1$ regular random graph, by generalizing the calculation of
the {\it non}-pinned system~\cite{shukla2008}.  The BP process
with random pinning is defined as follows:  (1) Prepare the $N$ sites
each of which is occupied with probability $p$.  (2) Pick up a faction
$c$ of sites randomly and ``pin'' them. (3) Recursively cull
$non-pinned$ sites that have less than $m$ occupied sites in neighbor.
Below, we focus on the case when $k=3$ and $m=3$ since it corresponds to
the model investigated in the main text, see
Ref.~\cite{sellitto2005} for more details about the
connection between the FA model and BP.

\subsection{ Scaling function of the probability of the occupied sites} First
we derive the scaling function for $P$ which is the fraction of the
occupied sites after the BP process.  To this end, it is more convenient
to consider the probability $P_*$ that a occupied site is not
culled in the BP process, given that one of the nearest neighbors
was occupied. $P_*$ follows the same scaling low of that of $P$, but
is easier to calculate~\cite{sellitto2005}.  $P_*$ satisfies the
following self-consistent equation~\cite{ikeda2015}:
\begin{align}
P_* = pc + p(1-c)\left[
3P_*^2 -2P_*^3 \right]. \label{182154_20May15}
\end{align}
It is convenient to introduce the auxiliary function defined by
\begin{align}
 Q(p,c,P_*) = cp + (1-c)p\left[3P_*^2-2P_*^3\right] - P_*.\label{111256_21May15}
\end{align}
At the $A_3$ transition point, $Q$ should satisfy following
equations~\cite{sellitto2013}:
\begin{align}
 Q &= cp + (1-c)p\left[3P_*^2-2P_*^3\right] - P_* = 0,\new
 \pdiff{Q}{P_*} &= 6(1-c)p(P_{*}-P_{*}^2)-1 = 0,\new
 \pdiff{^2Q}{P_*^2} &= 6(1-c)p(1-2P_{*}) = 0.
\end{align}
Solving the above equations, we obtain $c=1/5$, $p=5/6$ and $P_*=1/2$.
Expanding $Q$ around this point, one obtains
\begin{align}
 Q &= -\frac{4}{3}\delta P_*^3 + \frac{5}{12}\delta c
 + \frac{3}{5}\delta p -\frac{5}{4}\delta P_*\delta c + \frac{6}{5}\delta P_*\delta p +\ldots\label{113238_21May15}
\end{align}
where $\delta c=c-1/5$, $\delta p=p-5/6$ and $\delta P_*=P_*-1/2$.  
We decompose vector $(\delta c,\delta p)$ as
\begin{align}
 (\delta c, \delta p) = 
A\varepsilon_{\perp}\vec{e}_{\perp}
 +B\varepsilon_{\parallel}\vec{e}_{\parallel},\label{113231_21May15}
\end{align}
where $A$ and $B$ are arbitrary constants.  
The vectors, $\vec{e}_{\perp}$ and $\vec{e}_{\parallel}$, are defined as
 \begin{align}
  \vec{e}_{\perp} &= \left(\frac{5}{12},\frac{3}{5}\right), & \vec{e}_{\parallel} &= \left(-\frac{3}{5},\frac{5}{12}\right).
 \end{align}
Substituting
eq.~(\ref{113231_21May15}) into eq.~(\ref{113238_21May15}), we obtain
\begin{align}
 0 
&\sim  -\frac{4}{3}\delta P_*^3 + \frac{1921}{3600}A\varepsilon_{\perp} 
+ \frac{5}{4}B\delta P_* \varepsilon_{\parallel}.\label{114252_21May15}
\end{align}
Solving above equation, we obtain
\begin{align}
 \delta P*(\varepsilon_{\parallel},\varepsilon_{\perp}) = 
 \abs{\varepsilon_{\parallel}}^{\beta}
g\left(\frac{\varepsilon_{\perp}}{\abs{\varepsilon_{\parallel}}^{\delta\beta}}\right)
,\label{120008_21May15}
\end{align}
where $\beta=1/2$ and $\delta=3$.  The scaling function $g(y)$ is the
solution of
\begin{align}
 0 = g^3-\frac{1921}{4800}Ay\mp\frac{15}{16}Bg,\label{scaling}
\end{align}
where $\pm$ refers to the sign of $\varepsilon_{\parallel}$.  To clarify
the connection between this model and the mean-field random-field Ising
model, we set $A=\frac{4800}{1921}\times
\frac{12\sqrt{2}}{\pi^{3/2}R_c}$ and $B=\frac{16}{15}\times
\frac{12}{\pi}$.  Then eq.~(\ref{scaling}) is rewritten as
\begin{align}
 0 = g^3\mp\frac{12}{\pi}g-\frac{12\sqrt{2}}{\pi^{3/2}R_C}y = 0.\label{212314_24May15}
\end{align}
Above equation is identical to the scaling function of the order
parameter of the mean-field random-field Ising model (see eq.~(A.5) in
Ref.~\cite{PhysRevB.53.14872}).

\subsection{ Scaling function of the avalanche distribution} After the BP
process, all remained occupied sites which have more than $m$ neighbors
of occupied sites are ``blocked''.  Now we randomly cull a occupied
site. This leads the avalanche since some neighbors of the
culled site become unblocked and so on and so forth.  Here we consider
the distribution function (denoted as $\pi_a$) of the size of the
avalanche in a sub-tree connected to the culled site (see
Ref.~\cite{shukla2008} for more precise definition).  To this
end, it is convenient to introduce the generating
function\cite{shukla2008}:
\begin{align}
 \pi(x) = \sum_{a=0}^{\infty}\pi_ax^a.
\end{align}
$\pi(x)$ satisfies the self-consistent equation:
\begin{align}
 \pi(x) &= 
 (1-c)xp\sum_{k=0}^{z-1}\binom{z-1}{k}[\pi(x)]^k
\left[1-P_*\right]^{z-1-k}\delta_{k+1,m}\new
 &+\pi_0,\label{011929_15May15}
\end{align}
where 
\begin{align}
 \pi_0 = cp + (1-c)p\sum_{k=0}^{z-1}\binom{z-1}{k}[P_*]^k \left[1-P_*\right]^{z-1-k}p_k
\end{align}
is the probability that the avalanche does not occur. For $z=4$ and
$m=3$, the equation is
\begin{align}
 \pi(x) = cp+ (1-c)pP_*^3 + 3(1-c)xp(1-P_*)\pi(x)^2.\label{013450_15May15}
\end{align}
To investigate the avalanche distribution for the large $a$, we
substitute $x=1-\delta x$ and $\pi(x)=P_*+\delta\pi(x)$ into
eq.(\ref{013450_15May15}) and obtain
\begin{align}
 \delta\pi^2 +  C_1t\delta\pi -C_2\delta x = 0,
\end{align}
where $C_1$ and $C_2$ are constants, and we defined
\begin{align}
t  &
\sim
-4\abs{\varepsilon_{\parallel}}^{2\beta}\left[
g(\varepsilon_{\perp}/\abs{\varepsilon_{\parallel}}^{\delta\beta})^2 \mp\frac{5}{16}B
\right].
\label{122658_15May15}
\end{align}
Solving the above equation, we obtain
\begin{align}
 \delta\pi(x) = \frac{1}{2}\left[
-C_1t+\sqrt{C_1^2t^2+4C_2\delta x}
\right].
\end{align}
The asymptotic expression of the $\pi_a$ for large $a$ is given by
\begin{align}
 \pi_a = \frac{1}{a!}\left.\diff{^a\pi(x)}{x^a}\right|_{x=0}
\sim \frac{1}{a^{3/2}}e^{-C_3t^2a},\label{122943_15May15}
\end{align}
where $C_3=C_1^2/4C_2$.  Substituting eq.(\ref{122658_15May15}) into
eq.~(\ref{122943_15May15}), one obtains 
\begin{align}
 \pi_a(\varepsilon_{\parallel},\varepsilon_{\perp}) &\sim
a^{-3/2}
e^{
 -16C_3 a\abs{\varepsilon_{\parallel}}^{4\beta}
\left(
g(\varepsilon_{\perp}/\varepsilon_{\parallel}^{\delta\beta})^2\mp \frac{5}{16}B
\right)^2},
\end{align}
where $\tau=3/2$ and $\sigma=1/4\beta=1/2$. To see the connection
with the RFIM, we set $B=\frac{16}{15}\times \frac{12}{\pi}$, and
obtain 
\begin{align}
 \pi_a(\varepsilon_{\perp},\varepsilon_{\parallel})
\sim a^{-3/2}e^{
-\frac{2^8}{\pi^2}C_3a\abs{\varepsilon_{\parallel}}^{4\beta}
\left(
1\mp\frac{\pi}{4}g(\varepsilon_{\perp}/\varepsilon_{\parallel}^{\delta\beta})^2
\right)}.
\end{align}
Further, introducing the rescaled avalance size $S$ by
\begin{align}
 a = \frac{\pi^2}{C_32^9}S,
\end{align}
we obtain
\begin{align}
 \pi_S(\varepsilon_{\perp},\varepsilon_{\parallel})
 =S^{-\tau}{\cal D}(S/\abs{\varepsilon_{\parallel}}^{-1/\sigma},
\varepsilon_{\perp}/\abs{\varepsilon_{\parallel}}^{\beta\delta}
),
\end{align}
with the critical exponents $\tau=3/2$, $\sigma=1/2$, $\beta\delta=3/2$,
and the scaling function
\begin{align}
 {\cal D}(x,y) = e^{
-x\left[
1\mp \frac{\pi}{4}g(y)^2
\right]^2/2
}.
\end{align}
This equation corresponds to the scaling function of the avalanche
distribution of the random field Ising model (see eq.~(A12) in
Ref.\cite{PhysRevB.53.14872}).  To derive eq.~(12) in the main text,
one should note that $\varepsilon_{\perp}\sim
\abs{\varepsilon_{\parallel}} \sim \abs{T-T_c}$ and
$\varepsilon_{\perp}/\abs{\varepsilon_{\parallel}}^{\beta\delta} \gg 1$
for general directions.  Using $g(y)\sim y^{1/3}\ (y\gg 1)$, one arrives
at the asymptotic form of the avalanche distribution function
\begin{align}
P(S)
 \sim \pi_S 
 \sim S^{-\tau}e^{-CS\abs{T-T_c}^{4/3}},
\end{align}
where $C$ is a constan.

\bibliographystyle{eplbib.bst} 
\bibliography{./reference}

\end{document}